\documentclass[letterpaper,10pt,conference]{ieeeconf}  

\IEEEoverridecommandlockouts        
\usepackage{color,graphicx} 
\usepackage{mathptmx} 
\usepackage{times} 
\usepackage{amsmath} 
\usepackage{amssymb}  

\def\ket#1{|#1\rangle}
\def\bra#1{\langle #1|}
\def\ip#1#2{\langle #1|#2 \rangle}

\renewcommand{\>}{\rangle}
\newcommand{\<}{\langle}
\newcommand{\diag}{\mbox{diag}}

\renewcommand{\in}{\mathrm{in}}
\newcommand{\out}{\mathrm{out}}

\newcommand{\kin}{|\mathrm{in}\>}
\newcommand{\kout}{|\mathrm{out}\>}

\graphicspath{{figures/}}

\title{\LARGE \bf
Time optimal information transfer in spintronics networks
}

\author{Frank C Langbein$^{1,*}$, Sophie Schirmer$^{2,*}$ and Edmond Jonckheere$^{3,**}$%
         \thanks{*Supported by the Ser Cymru NRN.}
         \thanks{**Supported by ARO MURI.}
         \thanks{$^1$ School of Computer Science \& Informatics,
         Cardiff University, Cardiff, CF24 3AA, UK,
         {\tt\small LangbeinFC@cf.ac.uk}.}
         \thanks{$^2$ Dept. of Physics, College of Science,
         Swansea University, Swansea, SA2 8PP, UK,
         {\tt\small sgs29@swan.ac.uk}.}
         \thanks{$^3$ Dept. of Electrical Engineering,
         Univ.\ of Southern California,
         Los Angeles, CA 90089, USA,
         {\tt \small jonckhee@usc.edu}.}
}

\begin{document}
\maketitle
\thispagestyle{empty}
\pagestyle{empty}

\begin{abstract}
Propagation of information encoded in spin degrees of freedom through
networks of coupled spins enables important applications in
spintronics and quantum information processing. We study control of
information propagation in networks of spin-$\tfrac{1}{2}$ particles
with uniform nearest neighbor couplings forming a ring with a single
excitation in the network as simple prototype of a router for
spin-based information. Specifically optimizing spatially distributed
potentials, which remain constant during information transfer,
simplifies the implementation of the routing scheme.  However, the
limited degrees of freedom makes finding a control that maximizes the
transfer probability in a short time difficult.  We show that the
structure of the eigenvalues and eigenvectors must fulfill a specific
condition to be able to maximize the transfer fidelity, and
demonstrate that a specific choice among the many potential structures
that fulfill this condition significantly improves the solutions found
by optimal control.
\end{abstract}

\section{INTRODUCTION}

Encoding information spin degrees of freedom has the potential to
revolutionize information technology via the development of novel
spintronic devices and possible future quantum information
processors~\cite{spintronics1,spintronics2}.  Utilizing information
encoded in spin degrees of freedom, however, requires efficient,
controlled on-chip transfer of spin-based information.  In some types
of spintronic devices, spin degrees of freedom are used in addition to
motional degrees of freedom of electrons, and information encoded in
the spin degrees of freedom can be transferred using conventional
currents.  In principle, however, information stored in spin states
can propagate through a network of coupled spins without charge
transport.  As propagation of spin-based information is governed by
quantum-mechanics and the Schr\"odinger equation, however, excitations
in a spin network propagate, disperse and refocus in a wave-like
manner, and controlling information transport is thus a quantum
control problem.  Without any means to control the propagation of
spin-based information in such networks information transport can be
slow and inefficient. Control can be utilized to optimize
transport in terms of maximizing transfer efficiency and
speed~\cite{QIP2015,ISCCSP2014}.  Here, we consider how to
control information propagation in a network of spins by optimizing
spatially distributed potentials, which remain constant during the
evolution, in contrast to dynamic control schemes, which require
dynamic modulation or fast switching of the control potentials.

\section{THEORY AND METHODS}

\subsection{Networks of coupled spins}

We restrict ourselves to spin-$\tfrac{1}{2}$ particles with two spin
states labeled $\ket{0}$ and $\ket{1}$ as a simple prototype device
for routing spin-based information.  The energies of the spin states
of spin $n$ differ by small amounts $\Delta_n$.  Nearby spins can
interact, e.g., by exchange coupling.  This leads to a model
Hamiltonian for a network of coupled spins of the form
\begin{equation}
 \label{eq:H}
  H_{\rm full} = \sum_{n=1}^N \Delta_n Z_n + \sum_{m\neq n}
                J_{mn} [X_m X_n + Y_m Y_n + \kappa Z_m Z_n],
\end{equation}
where $\kappa$ is a parameter depending on the type of coupling, e.g.,
$\kappa=1$ for isotropic Heisenberg coupling and $\kappa=0$ for pure
$XX$ coupling.  $X_n$, $Y_n$ and $Z_n$ are operators acting on the
$2^N$ dimensional Hilbert space of the $N$-spin network.  $X_n$ is a
tensor product of $N-1$ identity operators with a single $X$ operator
in the $n$th position, and similarly for $Y_n$ and $Z_n$.  $X$, $Y$
and $Z$ are the Pauli spin operators
\begin{equation}
  X = \begin{pmatrix} 0 & 1  \\ 1 & 0 \end{pmatrix}, \quad
  Y = \begin{pmatrix} 0 & -i \\ i & 0 \end{pmatrix}, \quad
  Z = \begin{pmatrix} 1 & 0  \\ 0 & -1 \end{pmatrix}.
\end{equation}
For networks with uniform coupling all non-zero couplings have the
same strength $J$ and we can set $J=1$ by choosing the frequencies in
units of $J$ and time in units of $J^{-1}$.

It can be easily verified that a Hamiltonian of the form~(\ref{eq:H})
commutes with the total excitation operator $\sum_n Z_n$.  The Hilbert
space of the system therefore decomposes into excitation
subspaces~\cite{TAC2012}.  If we assume that only a single excitation
(or bit of information) propagates through the network at any given
time, then the Hamiltonian can be reduced to the single excitation
subspace Hamiltonian
\begin{equation}
  H = \Delta_n \ket{n}\bra{n} + J_{mn} \ket{m}\bra{n},
\end{equation}
where $\ket{m}\bra{n}$ can be thought of as a matrix which is zero
except for a $1$ in the $(m,n)$ position.  Which $J_{mn}$ are non-zero
depends on the network topology.  For a chain with nearest-neighbor
coupling we have $J_{mn}=0$ unless $m=n\pm 1$, and similarly for a
ring arrangement, except that for the latter we also have
$J_{N,1}=J_{1,N} \neq 0$.  While a linear chain can be thought of as a
type of quantum wire, a ring can be regarded as a basic routing
element to distribute spin states, e.g., to chains attached to various
nodes of the ring (see Fig.~\ref{fig:landscape}).

\subsection{Dynamic vs static control}

\begin{figure}
\includegraphics[width=\columnwidth]{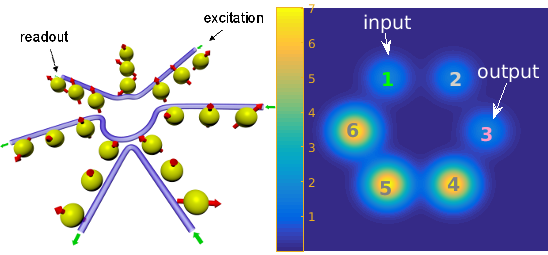}
\caption{Spin network ring extended with chains (left) and potential landscape
    created by local biases optimized for information transfer from
    node 1 to 3 in a 6 ring (right).}
\label{fig:landscape}
\end{figure}

Dynamic control involves dynamically altering certain couplings
$J_{mn}$ or potentials $\Delta_n$.  This is a powerful tool and has
been explored in previous work~\cite{PRA2009}.  However, it typically
requires the ability to rapidly modulate or switch
fields. Furthermore, for networks with a high degree of symmetry such
as rings with uniform coupling, controllability is generally limited
by dynamic symmetries~\cite{TAC2012}.  An alternative to this dynamic
control is to shape the potential landscape to facilitate the flow of
information from an initial state (input node) to a target state
(output node).  For example, for information encoded in nuclear spins
or electron spins in quantum dots whose potential can be controlled by
surface control electrodes, this could be achieved by varying the
voltages to create a potential landscape as shown in
Fig.~\ref{fig:landscape}.

Fixing the network topology defined by the couplings $J_{mn}$ and
applying potentials $\Delta_n$ that are constant in time, gives the
constant Hamiltonian $H_\Delta = H_0 + \diag(\Delta_n)$ and the
information transfer is governed by the Schr\"odinger equation
\begin{equation}
  i\hbar \dot{U}(t) = H_\Delta U(t), \quad U(0)=I,
\end{equation}
where $U(t)$ is a unitary propagation operator, initially equal to the
identity operator $I$. The probability of transmission of information
from the input node $\ket{\in}$ to the output node $\ket{\out}$ in
time $t$ is then given by
\begin{equation}
   p(t) = |\bra{\out} e^{-i tH_\Delta}\ket{\in}|^2.
\end{equation}
Assuming the energies $\Delta_n$ are controllable, we have a vector of
control parameters $\Delta = (\Delta_n)$ and the objective is to find
a $\Delta$ such that
\begin{equation}
  p(t) = \max_{\Delta} |\bra{\out} e^{-i tH_\Delta} \ket{\in}|^2
  \label{eq:opt1}
\end{equation}
at some time $t$.  We can fix the time $t=t_f$, require that $t\le
t_{\max}$ where $t_{\max}$ is an upper bound, or attempt to accomplish
the transfer with maximum fidelity in minimum time.  These problems
can in principle be solved in a straightforward manner using standard
optimization tools although the optimization landscape is challenging,
in particular when the goal is to find a control that achieves the
highest possible fidelity in the shortest time possible as this
effectively creates two objectives for the optimization.

\begin{figure*}\centering
\includegraphics[width=.98\textwidth]{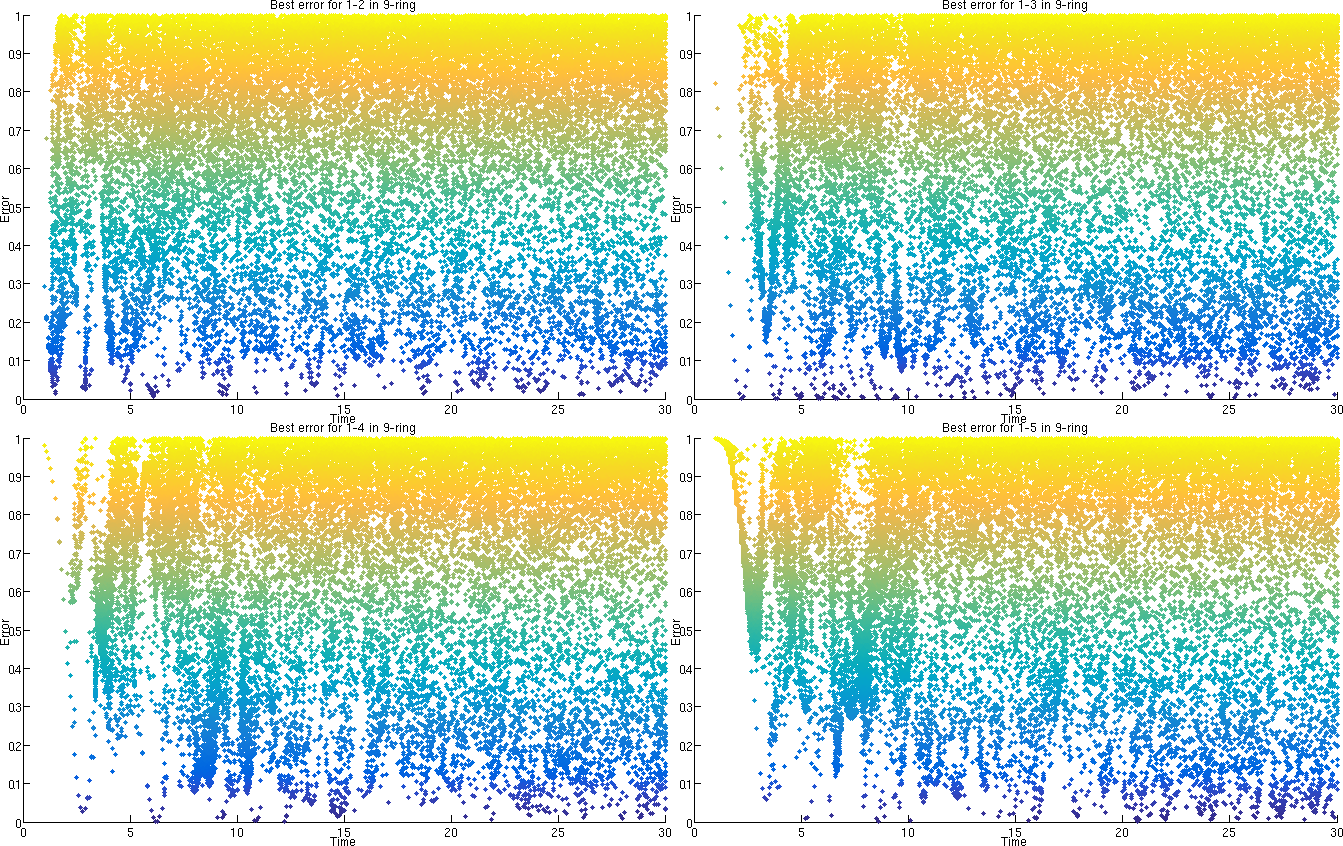}
\caption{Results of optimizing the information propagation from spin
  $1$ to $2$ (top left), $3$ (top right), $4$ (bottom left) and $5$
  (bottom right) for a ring of $9$ spins using a quasi-Newton L-BFGS
  optimization with exact gradient on static biases for fixed times
  $f_f$ from $1$ to $30$ with step size $0.2$. Each data point
  represents the infidelity $1-p(t)$ achieved for the corresponding
  time, indicating that the optimization gets trapped often, even if
  many repeats can still find good solutions for certain times.}
\label{fig:ring9-errors}
\end{figure*}

\subsection{Eigenstructure optimization} \label{sec:eigen}

We can reformulate the control problem by diagonalizing the
Hamiltonian, $H_\Delta = V \Lambda V^\dag$, where $V$ is a unitary
matrix of the eigenvectors and $\Lambda$ a diagonal matrix of the
eigenvalues of $H_\Delta$.  Then $U = e^{-itH_\Delta} = VE V^\dag$
with $E=\diag(e^{-i\lambda_nt})$ and our control objective is to
ensure
\begin{equation}
\label{eq:1}
   V^\dag\ket{\out} = e^{i\phi} E V^\dag \ket{\in},
\end{equation}
where $\phi$ is a global phase factor, which we introduce to cancel
the the global phase of the output state.

For a network with ring topology, we can take the input state to be
$\ket{1}=(1,0,\ldots)^T$ due to translation invariance, and the
output state to be $\ket{k}$, $k \le \lceil N/2 \rceil$. Then
Eq.~\eqref{eq:1} becomes
\begin{equation}
\label{eq:2}
   v_{n,k}^* = e^{-i(t \lambda_n -\phi)} v_{n,1}^*, \quad \forall n=1,\ldots, N.
\end{equation}
Thus, the optimization problem is equivalent to finding the a control
vector $\Delta$ and phase $\phi$ that minimize
\begin{equation}
\label{eq:3}
  \sum_n |v_{k,n} - e^{i(t \lambda_n-\phi)} v_{1,n}|^2
\end{equation}
Note that only the $1$st and $k$th component of the eigenvectors (or
$1$st and $k$th rows of $V$) and eigenvalues $\lambda_n$ matter.

The upper bound on the fidelity of the transmission given no time
constraints, referred to as Information Transfer Fidelity~\cite{ISCCSP2012,QIP2014}, is given by
\begin{equation}
\label{e:ITF}
  \sqrt{\mathrm{ITF}} = \sum_{n=1}^N  |\<\in | v_n\>\<v_n | \out\>|
=\sum_{n=1}^N  s_n\<\in | v_n\>\<v_n | \out\>,
\end{equation}
where $s_n=\mathrm{sign}(\<\in | v_n\>\<v_n | \out\>)$. Since it is a
probability, $\mathrm{ITF} \leq 1$. Here we use some eigenstructure
assignment concept~\cite{eigenstructure} to show that, given $\kin$
and $\kout$, there always exists an orthonormal reference frame
$\{v_n:n=1,...,N\}$ such that the upper bound $\mathrm{ITF} \leq 1$ is
achieved.  The issue as to whether this reference frame can be
achieved by an optimal biasing strategy $\Delta = (\Delta_n)_{n=1}^N$
is open, but numerical exploration seems to indicate that the optimal
biases in the sense of Sec.~\ref{s:results} reproduce this frame with
high accuracy.

Instead of deriving the position of the reference frame
$\{v_k:k=1,...,N\}$ relative to the input and output states, we
reformulate the problem as the one of finding the position of the
input and output states relative to a given (orthonormal) reference
frame $\{v_n:n=1,...,N\}$.  Let $\ip{v_n}{\in}$, $\ip{v_n}{\out}$,
$n=1,...,N$, be the coordinates of the input and output states, resp.,
in the reference frame $\{v_k:k=1,...,N\}$. The problem is to
maximize Eq.~(\ref{e:ITF}) subject to the constraints
\[
  \ip{\out}{\in} =0, \quad  \ip{\in}{\in}=1, \quad \ip{\out}{\out}=1.
\]
Define the Lagrange multipliers $\lambda$, $\mu$, $\kappa$ and the augmented functional
\[
 \sum_{n=1}^N s_n \ip{v_n}{\in}^* \ip{v_n}{\out} +\lambda \ip{\out}{\in}
  +\tfrac{\kappa}{2} \ip{\in}{\in} +\tfrac{\mu}{2} \ip{\out}{\out}
\]
The classical first order condition for optimality yields
\begin{equation}
\label{e:first_order}
\begin{pmatrix} \lambda+s_n & \mu\\ \kappa & \lambda +s_n \end{pmatrix}
\begin{pmatrix} \ip{v_n}{\in} \\ \ip{v_n}{\out} \end{pmatrix} = 0,
\quad n=1,...,N.
\end{equation}
Existence of a solution yields
\[
  \lambda= (\pm) \sqrt{\kappa \mu} - s_n.
\]
The {\it crucial issue} is to observe that $\lambda$ should be
independent of $n$.  At optimality, not all $s_n$'s could be of the
same sign (otherwise $\mathrm{ITF}=|\ip{\out}{\in}|^2=0$).  This
implies that $s_n=(\pm)$ and $\sqrt{\kappa \mu}=1$.  From there on,
solving~\eqref{e:first_order}, and after some manipulation, it is
found that
\[
  \ip{v_n}{\in} =\pm \ip{v_n}{\out}, \quad n=1, \ldots ,N.
\]
As already said, the optimal biases $\Delta_n$'s appear to reproduce
this result. The above eigenvector assignment is non-unique and
dimension dependent. When $N=2$, the solution is already far from
unique. Up to permutation, either $\{\kin,\kout\}$ is in
$\mathrm{span}\{v_1,v_2\}$ and offset at a $45^{\circ}$ angle or
$\kin=(1/\sqrt{4})(1,1,1,1)$ and $\kout=1/\sqrt{4}(1,-1,1,-1)$.


\section{RESULTS}
\label{s:results}

\subsection{General optimization results}

Solving the optimization problem given by Eq.~(\ref{eq:opt1}) directly
for a fixed target time $t_f$ is challenging even without constraints
on the biases $\Delta$ as the landscape is extremely complicated with
many local extrema, resulting in trapping of local optimization
approaches such as quasi-Newton methods.  Fig.~\ref{fig:ring9-errors}
shows the results of various runs for fixed times for a ring of $9$
spins, with the objective being to propagate the excitation from spin
$1$ to spin $2$, $3$, $4$ and $5$, respectively.  Fixed time values
$t_f$ from $1$ to $30$ with steps of $0.2$ were set and a quasi-Newton
optimizer with random initial values for the biases repeated $100$
times for each $t_f$. Each point in this plot represents one run for
the respective time $t_f$ and the achieved infidelity value $1-p(t)$.
There are many points for which the error is large (i.e.,
the transmission fidelity is low), indicating that the optimization
for this run converged to a local extremum, making finding a good
solution very expensive.  We also tested some global search
algorithms that appeared successful for hard optimization problems
such as a variant of differential evolution used in~\cite{PRA2015},
but these proved to be much slower and produced generally worse
results than the best solutions found by repeating a quasi-Newton
optimization algorithm for a relatively large number of initial
states.  The results in Fig.~\ref{fig:ring9-errors} also indicate that
there are only certain times for which we can expect to find a high
information transfer fidelity.

\begin{figure}
\includegraphics[width=\columnwidth]{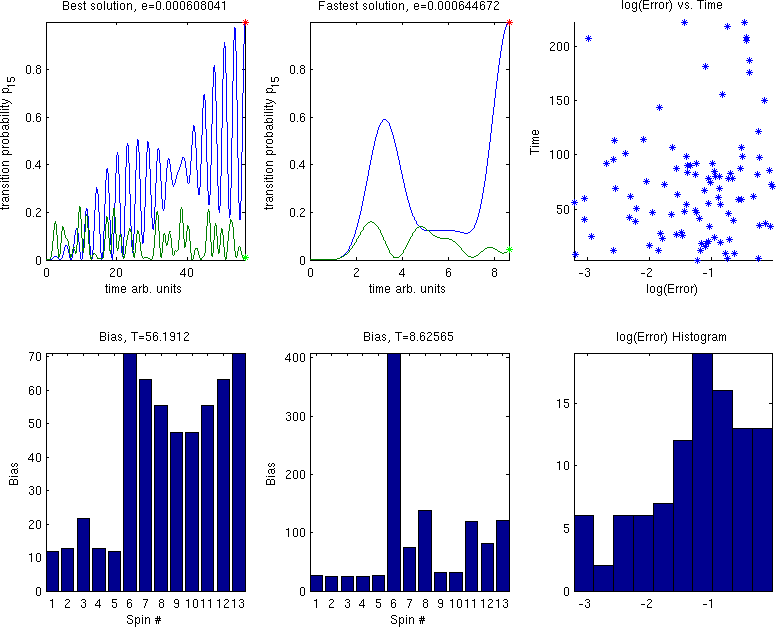}
\caption{Results for optimizing the information transfer probability
  from spin $1$ to $5$ for a ring of $13$ spins over the spatial
  biases and times starting from random initial biases and
  times. Left-bottom shows the biases giving the best fidelity at time
  $T \approx 56.19$ yielding the $p(t)$ shown in the left-top image
  (in blue vs. the natural evolution in green). The two graphs in the
  middle show the fastest solution found with a fidelity greater than
  $0.99$ at time $T \approx 8.63$. Right-top shows the overall found
  solutions by the optimization plotting the time vs. the logarithm of
  the infidelity. Right-bottom plots a histogram of the logarithm of
  the infidelity.}
\label{fig:ring13_1-5}
\end{figure}

Instead of fixing the time we can also add the time as additional
parameter to optimize over and start for random initial biases and
times. The results of this are shown in Fig.~\ref{fig:ring13_1-5} for
the propagation from spin $1$ to $5$ in a ring of $13$ spins. We
report the solution with the highest fidelity overall and the fastest
solution with a fidelity larger than $0.99$. Typically the highest
fidelity solutions are found at longer times but good solutions for
short times are also achieved.  However, many restarts of the
optimization are required, and many runs fail with fidelities smaller
than $0.9$.  Inspection of the good solutions found showed that many
of these involved biases that were highly mirror symmetric w.r.t. the
symmetry axis between the input and output spin on the ring.

\subsection{Finding good initial values}

In principle the information transfer fidelity depends on all
eigenvectors and eigenvalues of the Hamiltonian. However, the results
in Section~\ref{sec:eigen} show that the structure of the eigenvalues
and eigenvectors must fulfill a specific condition to be able to
maximize the transfer fidelity.  While there are many potential
structures that fulfill the condition, we can choose a specific one to
provide a guide for good initial values or a restricted domain for the
population.  This significantly improved the solutions found by the
optimal control algorithms.

\begin{figure}
\includegraphics[width=\columnwidth]{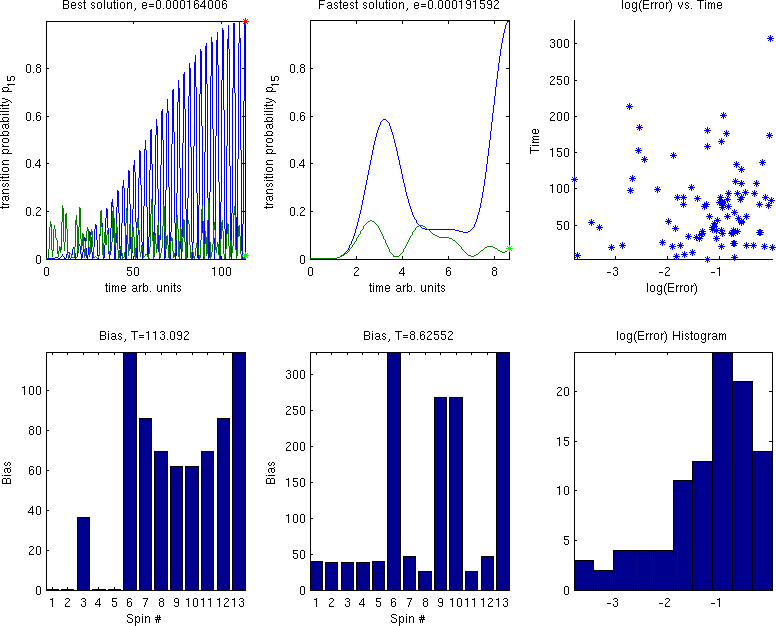}
\caption{Results for optimizing the information transfer probability
  from spin $1$ to $5$ for a ring of $13$ spins over the spatial
  biases and times starting from random initial biases and times,
  where the biases are constrained to be symmetric across the axis
  going through the middle between initial and target
  state. Left-bottom shows the biases giving the best fidelity at time
  $T \approx 113.09$ yielding the $p(t)$ shown in the left-top image
  (in blue vs. the natural evolution in green). The two graphs in the
  middle show the fastest solution found with a fidelity greater than
  $0.99$ at time $T \approx 8.63$. Right-top shows the overall found
  solutions by the optimization plotting the time vs. the logarithm of
  the infidelity. Right-bottom plots a histogram of the logarithm of
  the infidelity.}
\label{fig:ring13_1-5_symmetry}
\end{figure}

The basic idea for imposing a specific eigenstructure is to quench the
ring of $N$ spins into a chain from the initial spin to the target
spin. Our previous work showed that this can be easily achieved by
applying a very strong potential in the middle between the initial and
target spin~\cite{QIP2015}.  If we can control the potentials of all
spins then we can generalize this to quench the ring just before the
initial and after the target node, giving two options for a chain
connecting the two nodes where either could provide a solution. As it
turns out, this gives rise to a general eigenstructure induced by
applying mirror symmetric potentials across the axis going through the
middle between initial and target state in the ring.  We consequently
choose such symmetric potentials in combination with the approximate
times where the maximum fidelity is achieved in the related chains as
initial values for the optimal control algorithm. This significantly
improves the results and efficiently finds controls for maximum
information transfer in minimum time for any initial and target
spin. This approach is also further justified by the results found
using random initial values.

\begin{figure}
\includegraphics[width=\columnwidth]{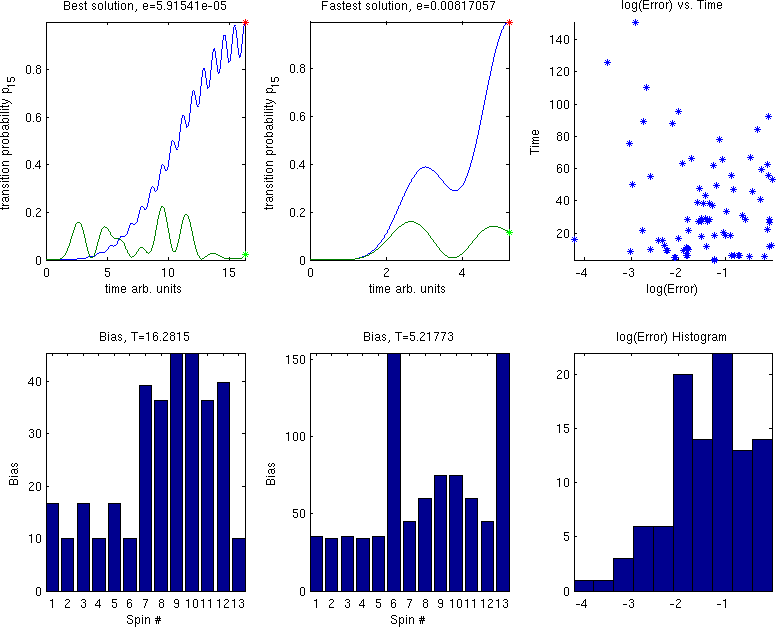}
\caption{Results for information transfer from spin $1$ to
    $5$ in a ring of $13$ spins, optimizing both spatial biases and
    transfer times, starting with initial times corresponding to peaks
    in the chain transition and random biases.  Left-bottom shows the
    biases giving the best fidelity at time $T \approx 16.28$ yielding
    the $p(t)$ in left-top graph (in blue vs. the natural evolution in
    green). The graphs in the middle show the fastest solution found
    with a fidelity greater than $0.99$ at time $T \approx
    5.22$. Top-right plot shows the time-error distribution of the
    solutions found over $100$ runs and bottom right graphs shows a
    histogram of the logarithmic infidelity.}
\label{fig:ring13_1-5_time}
\end{figure}

\begin{figure}
\includegraphics[width=\columnwidth]{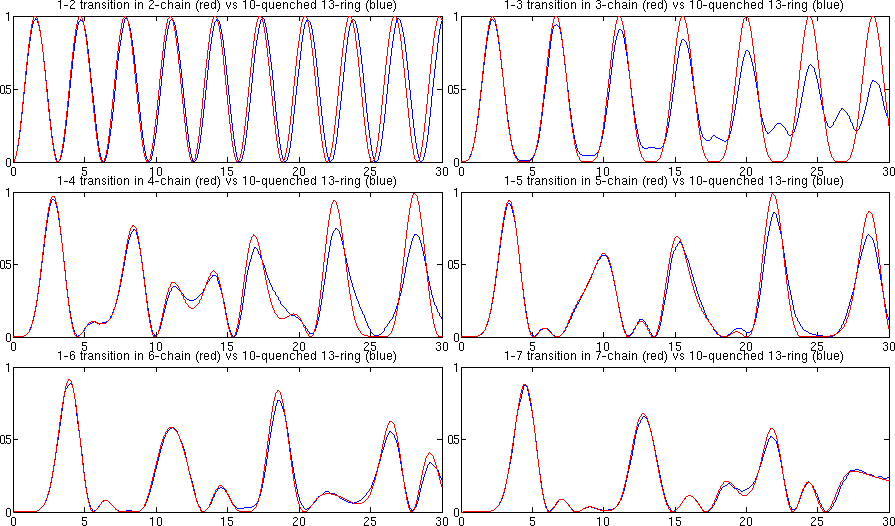}
\caption{Information transfer probabilities for $1$ to $k=2,3,4,5,6,7$
  in $13$-ring (blue) with a bias of $10$ on the spins from $k+1$ to
  $13$ vs. end-to-end natural transfer in $k$ chain (red)}
\label{fig:ring_v_chain}
\end{figure}

\begin{figure}
\includegraphics[width=\columnwidth]{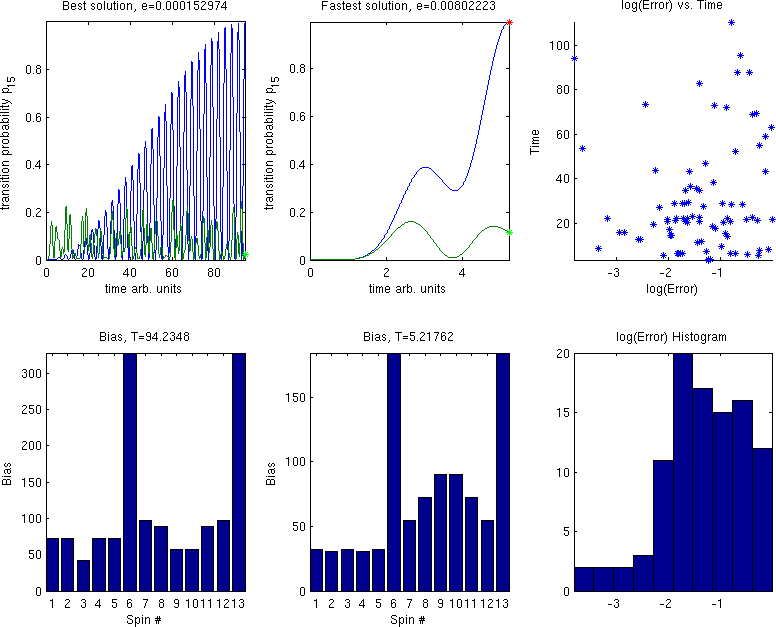}
\caption{Results for information transfer from spin $1$ to $5$ for a
  ring of $13$ spins, optimizing both transfer times and spatial
  biases, starting with initial times derived from the peaks of the
  corresponding chain transition and biases that are random but
  subject to symmetry constraints. Left-bottom shows the biases giving
  the best fidelity at time $T \approx 94.23$ yielding the $p(t)$
  shown in the left-top image (in blue vs. the natural evolution in
  green). Middle graphs show the fastest solution found with a
  fidelity greater than $0.99$ at time $T \approx 5.22$. Right-top
  shows the overall found solutions by the optimization plotting the
  time vs. the logarithm of the infidelity. Right-bottom plot shows
  histogram of logarithmic infidelity.}
\label{fig:ring13_1-5_symmetry_time}
\end{figure}

The symmetry constraint can be easily applied by reducing the number
of biases to be found to $\lceil N/2\rceil$, symmetric across the
symmetry axis between initial and target state.
Fig.~\ref{fig:ring13_1-5_symmetry} shows the results with this
constraint similar to the results in Fig.~\ref{fig:ring13_1-5} without
the symmetry constraint. Similar shortest time solutions are found,
while the best solution is at a different time. This is not surprising
considering that the solution found strongly depends on initial
values. More importantly, there are slightly fewer failed runs.
Convergence of the optimization can be further improved by selecting
constants, peaks or troughs as biases between initial and target spin
on both sides of the ring randomly.

Fig.~\ref{fig:ring_v_chain} compares the information transfer
probabilities from spin $1$ to $k$ in an $N$ ring with a constant bias
on the spins $k$ to $N$ with the information transfer probabilities
between the end nodes of a chain of length $k$.  From this example and
theory~\cite{QIP2015} it is obvious that the maxima
coincide and the stronger the bias the more similar the information
transfer probabilities between the quenched ring and the chain. Hence,
it seems likely that we can obtain better results if the initial times
are taken from the largest peaks, say those over $0.8$, of the chain
transition, which can easily be approximated by evaluating the
probabilities for the chain at a regular sampling.

\addtolength{\textheight}{-1.5cm}   

Fig.~\ref{fig:ring13_1-5_time} shows the results for the $1$ to $5$
transition of the $13$ ring for random, unconstrained initial biases and
the initial times taken from the peaks of the corresponding $1$ to $5$ transition
in a $5$ chain. This resulted in a faster time being found, but still
quite a few failed runs. Combining this with the symmetry constraint
gives the results shown in Fig.~\ref{fig:ring13_1-5_symmetry_time},
resulting in the same short time and fewer failed runs.

\begin{figure}
\includegraphics[width=\columnwidth]{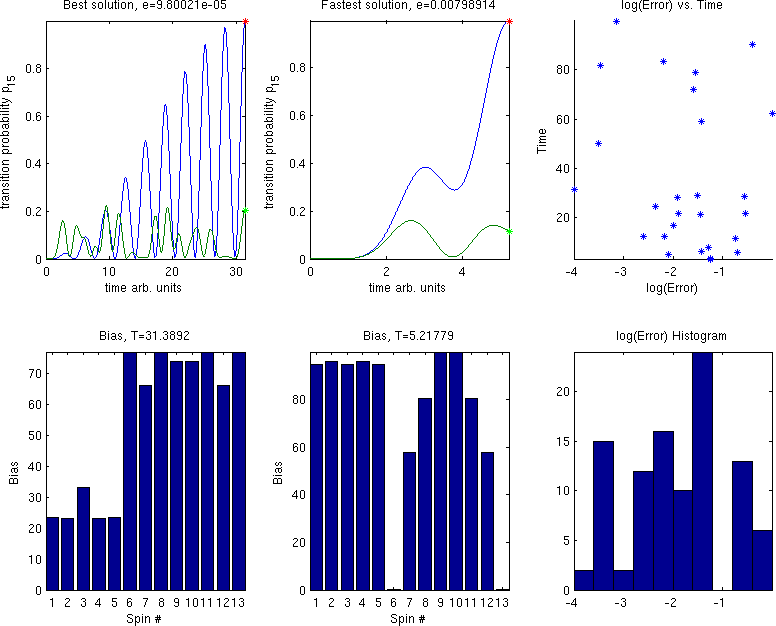}
\caption{Results for information transfer from spin $1$ to $5$ in a
  ring of $13$ spins, optimizing both spatial biases and transfer
  times, starting with initial times corresponding to peaks in the
  chain transition and biases that are random but subject to symmetry
  and amplitude constraints $0\le \Delta_n \le 100$. Left-bottom shows
  biases giving the best fidelity at time $T \approx 94.23$ yielding
  $p(t)$ shown in the left-top image (in blue vs. natural evolution in
  green).  The graphs in the middle show the fastest solution found
  with a fidelity $>0.99$ at time $T \approx 5.22$. Top-right plot
  shows the time-error distribution of the solutions found over $100$
  runs and bottom right graphs shows a histogram of the logarithmic
  infidelity.}
\label{fig:ring13_1-5_cons}
\end{figure}

We can further add a constraint to limit the strength of the
biases. Fig.~\ref{fig:ring13_1-5_cons} shows the results. To achieve
the same effect of quenching the ring to a chain, all potentials are
moved towards the maximum instead of pushing the two potentials at the
two end spins to very high values.

\begin{figure}
  \centering
\includegraphics[width=0.9\columnwidth]{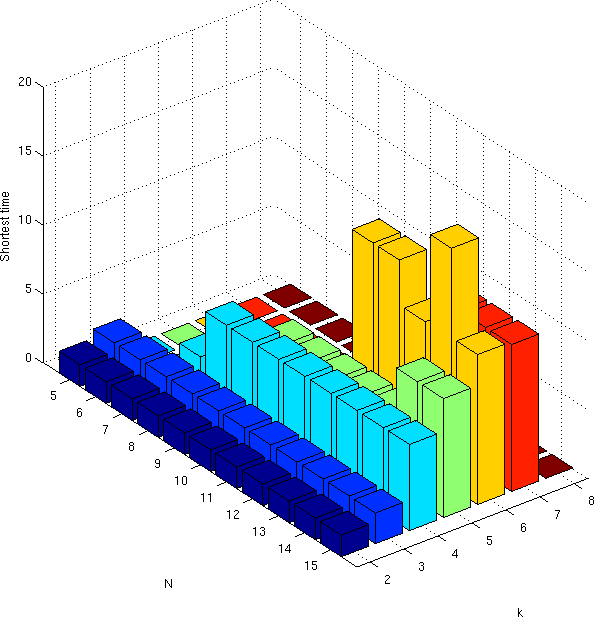}
\caption{Shortest times achieved for transition fidelities greater
  than $0.99$ for rings of size $N=5,\dots,15$ and transitions from
  $1$ to $k=2,\dotsc,\lceil N/2 \rceil$}
\label{fig:shortest}
\end{figure}

Overall, results for other rings and transitions are similar to the
results presented for the $1$ to $5$ transition for a ring of size
$N=13$.  Fig~\ref{fig:shortest} shows the shortest times achieved for
transition fidelities greater than $0.99$. This indicates that the
shortest transition times depend largely on the distance between the
spins and not the size of the ring, consistent with quenching the ring
into a chain.  Although we have no proof that there is no shorter
transition time, the results seem to indicate that the shortest time
in the ring is closely related to the time of the natural evolution to
the corresponding chain, in case of static bias controls.

In special cases we can further derive values for the expected minimum
transfer time and the corresponding biases.  If the distance between
the initial and target spin is $1$ then quenching the ring as
described reduces the network to a two-spin system with direct
coupling and an effective Hamiltonian of the form $H = \begin{pmatrix}
  c_1 & 1 \\ 1 & c_2 \end{pmatrix}$ which undergoes Rabi oscillations
with the Rabi frequency $\Omega=\sqrt{(c_2-c_1)^2+4}$.  It can easily
be shown that
$p_{12}(t)=(\tfrac{1}{2}\Omega)^{-2}\sin^2(\tfrac{1}{2}{\Omega}t)$. The
maximum $p_{12}=1$ is assumed for $t=\tfrac{\pi}{2}$, if and only if
$\Omega=2$, or $c_1=c_2$.  Indeed the numerical optimization results
in Fig.~\ref{fig:ring9-errors} for transfer from $1$ to $2$ (top left)
in a ring of size $9$ show that the first minimum is close to $0$ and
occurs for $t\approx \pi/2$.

Similarly, if the distance is $2$ the ring is reduced to a three-spin
chain.  In this case we can easily show that, assuming zero-bias,
$\Delta=(0,0,0)$, $p_{13}=\sin^4(\tfrac{1}{2}\sqrt{2} t)$ and thus
$p_{14}=1$, i.e., we have perfect state transfer, for
$t=\tfrac{\pi}{2}\sqrt{2}$.  Fig.~\ref{fig:ring9-errors} (top-right)
shows that the numerical results are indeed consistent with this
solution with the first minimum for the $1\to3$ transfer occurring for
$t\approx\tfrac{\pi}{2}\sqrt{2}$.

More generally, if the distance between the input and output nodes is
$k-1$ and the biases satisfy $c_{k+1-n}=c_n$ then $H$ commutes with
the permutation $\sigma=[k,k-1,\ldots,1]$.  If $P=P^\dag$ is the
corresponding permutation matrix then $PHP = H$ and thus $V\Lambda
V^\dag = PV \Lambda V^\dag P$ or $V=PV$.  In particular, this means
that the first and last row of $V$ are the same and Eq.~(\ref{eq:3})
becomes
\begin{equation}
  \sum_n |v_{1,n}|^2 |1 - e^{i(t \lambda_n-\phi)}|^2
   = 4 \sum_n |v_{1,n}|^2  \sin^2(\tfrac{1}{2}(t \lambda_n-\phi)).
\end{equation}
This expression vanishes if $t\lambda_n -\phi$ is a multiple of $2\pi$
for $n=1,\ldots,k$.  In the previous example, for a chain of length $3$
with no bias, $\lambda_1=-\lambda_3=\sqrt{2}$ and $\lambda_2=0$, hence
we achieve perfect state transfer for
$t=2\pi/\lambda_1=\frac{1}{2}\sqrt{2}\pi$ setting $\phi=0$.


\section{CONCLUSIONS AND FUTURE WORK}

We have demonstrated how static controls can be used to control the
information flow in $\tfrac{1}{2}$-spin rings. Compared to
  dynamic control, finding static controls is a considerably harder
  optimization problem due to the complexity of the optimization
  landscape. Careful selection of initial values and enforcement of
  the constraints derived from eigenstructure analysis substantially
  improve the performance of the algorithm.  For rings, the
  constraints enforce symmetries that make the rings more similar to
  chains, and the timing of transmission peaks for corresponding
  chains give a good indication of the shortest possible times that
  can be achieved.  Better understanding of the symmetry constraints,
  their relation to the reachability of target states and the design of
  global optimization algorithms that utilize the specific structure
  of the problem may yield even better solutions.

\end{document}